\documentclass{appolb}
\usepackage{epsfig}
\usepackage{4_here}
\usepackage{cite}

\begin{document}
\title{ISOSPIN DEPENDENCE OF THE $\eta$ MESON PRODUCTION IN HADRONIC COLLISIONS%
\thanks{Presented at the Symposium on Meson Physics, Cracow, 01-04 October 2008.}%
}
\author{Rafa{\l}~Czy{\.z}ykiewicz$^{1}$ and Pawe{\l}~Moskal$^{1,2}$ \\
 on behalf  of the  COSY-11 collaboration
\address{
$^{1}$Institute of Physics, Jagellonian University, Cracow, Poland\\
$^{2}$Institute for Nuclear Physics and J{\"u}lich Center for Hadron Physics, \\
Research Center J{\"u}lich, Germany
}}
\maketitle
\begin{abstract}
Based on recent COSY-11 results of measurements of total cross sections for the quasi-free $pn\to pn\eta$ reaction
we determine the isospin I~=~0 component of the total cross section for the $NN\to NN\eta$ reaction down to the threshold. 
We show that the energy dependence of the total cross section ratios $\frac{\sigma_{I=0}(pn\to pn\eta)}{\sigma(pp\to pp\eta)}$ 
and $\frac{\sigma_{I=0}(pn\to pn\eta)}{\sigma(pn\to d\eta)}$ 
can be described using the F{\"a}ldt  and Wilkin analitical parametrization of the  nucleon-nucleon final state interaction.
\end{abstract}

\section{Introduction}

Studies on the $\eta$ meson production in hadronic collisions via different isospin channels have
had a large contribution to the understanding of the reaction mechanism~\cite{johansson,review,hanhart}.
From the comparison 
of the total cross sections for reactions $pn\to pn\eta$~\cite{calenpn,c11pneta,c11pneta2,moskalaip} 
and $pp\to pp\eta$~\cite{total1,total2,total3,total4,total5,total6,total7,total8} 
it was derived that the production of the $\eta$ meson with the total 
isospin $I~=~0$ in the initial channel exceeds the production with the 
isospin $I~=~1$ by over an order of magnitude, 
suggesting~\cite{wilkinratio} 
the isovector meson exchange 
to be the dominant process leading to 
excitation of the S$_{11}$ resonance. 
This mechanism is considered
to be predominant~\cite{theor1,theor2,theor3,theor4,theor5,theor6,theor7,wilkin,nakayama}.
However, relative contributions to the production process of  $\pi$ and  $\rho$ meson exchange
are still not well settled~\cite{nakayama5,shyam5,colin5,xie5}.

In this paper we determine contributions of the 
I~=~0 and I~=~1 components to the total cross section of  the $NN\to NN\eta$ 
reaction taking into account an entire available data base including our recent 
cross sections for the $pn\to pn\eta$ reaction  determined near the kinematical threshold~\cite{c11pneta}.

\section{$pn\to pn\eta$ and $pp\to pp\eta$ total cross section ratio}

\begin{figure}[H]
\begin{center}
  \includegraphics[width=6.2cm]{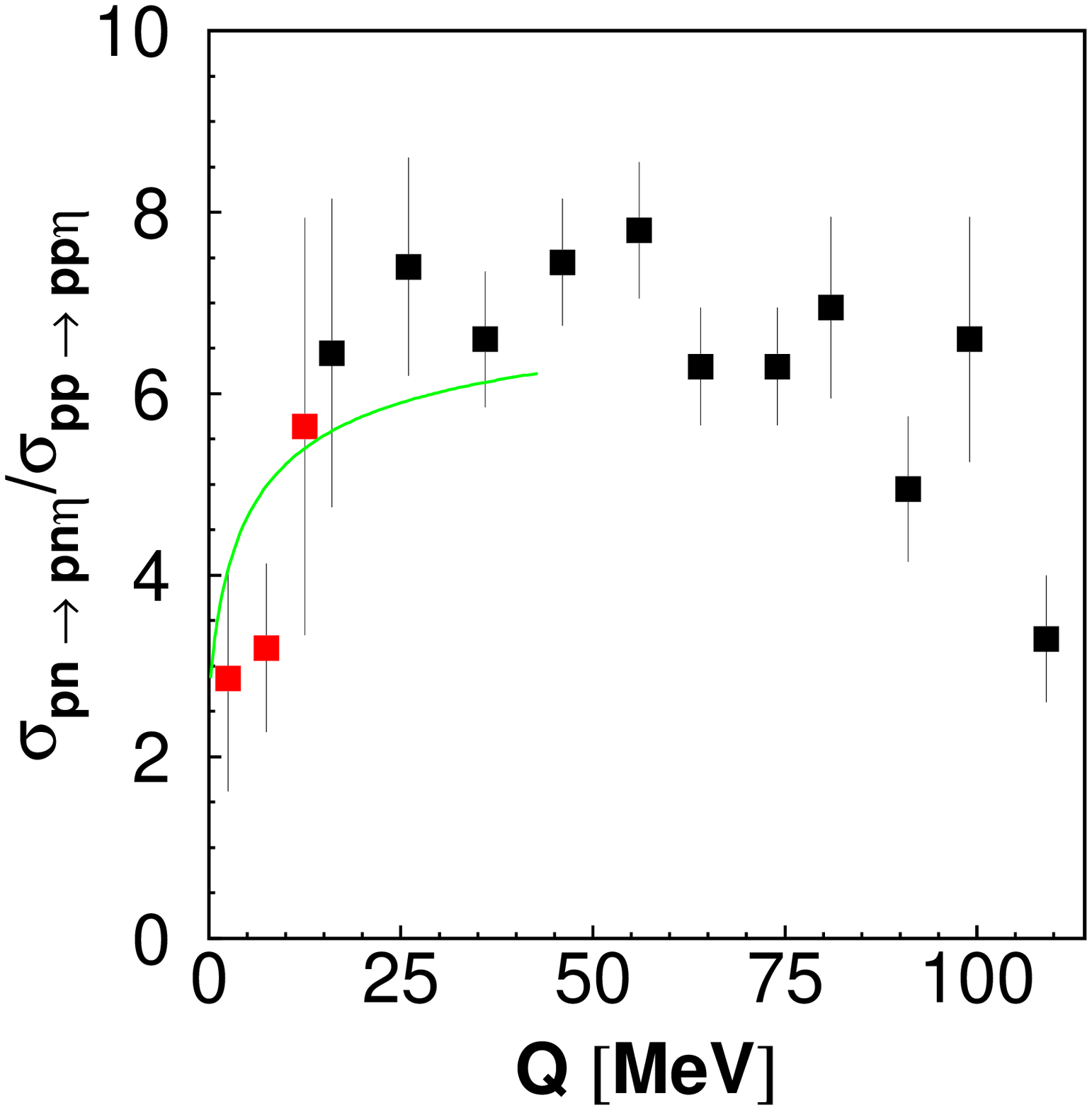}
  \includegraphics[width=6.1cm]{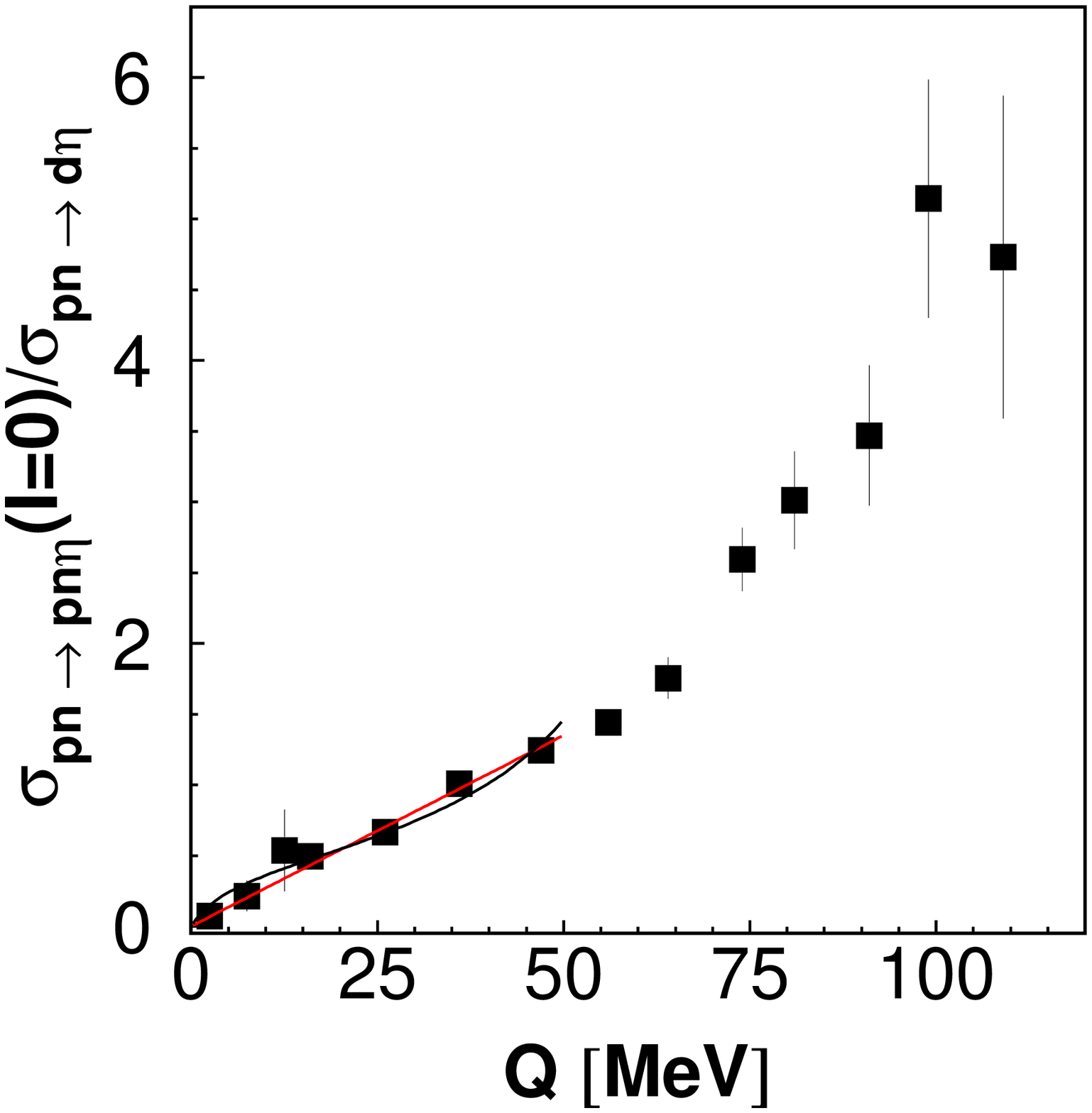}
\end{center}
\caption{ (left) Ratio of the total cross sections for the $pn\to pn\eta$ 
and $pp\to pp\eta$ reactions.
A superimposed line indicates a result of the fit taking into account
the final state interaction of nucleons.
(right) Ratio between the $I~=~0$ component of the $pn\to pn\eta$ total cross section and the total cross section for the $pn\to d\eta$. A superimposed lines indicate  result 
of  fits taking into account
the final state interaction of nucleons (black line), and assuming that 
the ratio of the total cross sections changes linearly with the excess energy (straight line). 
\label{pneta-deta}}
\end{figure}

Denoting by $\sigma_0$ and $\sigma_1$ the isospin I~=~0 and I~=~1 
components of the
total cross section for the $NN\to NN\eta$ reaction, 
we can write that:
\begin{equation}
\sigma(pn\to pn\eta) = \frac{1}{2}(\sigma_0 + \sigma_1),
\label{skladowe}
\end{equation}
and that 
\begin{equation}
\sigma(pp\to pp\eta) = \sigma_1.
\label{eq3}
\end{equation}
Left panel of Figure~\ref{pneta-deta} shows the ratio 
of the total cross sections
for the $pn\to pn\eta$ reaction to the total cross section for the $pp\to pp\eta$ reaction
plotted as a
function of the excess energy.
It was surprising to observe this ratio to fall down at lower values of~Q.
However, as explained by Wilkin~\cite{colinpriv2008}, to large extent, 
this behavior may  plausibly be assigned to 
the difference in strength of the proton-proton
and proton-neutron FSI.
Following the reference~\cite{fw1} the parameterization of the
isospin I~=~0 component of the cross section for the
$pn\to pn\eta$ reaction, taking into account proton-neutron FSI, is given by:
\begin{equation}
\sigma_0(pn\to pn\eta) = A \frac{Q^2}{(1+\sqrt{1+Q/\epsilon_{pn}}\ )^2},
\label{faldt}
\end{equation}
where $A$ is a constant, $Q$ is an excess energy, and $\epsilon_{pn}~=~2.2$~MeV 
is the binding energy of the $pn$ bound state~\cite{colinpriv2008}.

Analogously, the parameterization of the $pp\to pp\eta$ reaction total cross section
(pure isospin I~=~1) is given by: 
\begin{equation}
\sigma(pp\to pp\eta) = B \frac{Q^2}{(1+\sqrt{1+Q/\epsilon_{pp}}\ )^2},
\label{faldt2}
\end{equation}
with $\epsilon_{pp}~=~0.68$~MeV being the ''binding" energy of the 
pp virtual state~\cite{fw1}, and $B$ being a constant.
The value of $\epsilon_{pp}~=~0.68$ was derived~\cite{pawel2} from the 
fit of formula~\ref{faldt2} to the cross sections 
for the $pp\to pp\eta^{\prime}$ reaction~\cite{etap1,etap2,etap3,etap4}
for which the influence from the proton-meson final state interaction can be neglected~\cite{hab}.

Employing equations~\ref{faldt2}, \ref{faldt}, \ref{eq3}, and \ref{skladowe}
one obtains for the cross sections ratio a following closed analytical formula 
which accounts for the interaction between nucleons~\cite{fw1,fw2}:
\begin{equation}
\frac{\sigma(pn\to pn\eta)}{\sigma(pp\to pp\eta)} = 0.5 + C (\frac{\sqrt{\epsilon_{pp}}+\sqrt{\epsilon_{pp}+Q}}{\sqrt{\epsilon_{pn}}+\sqrt
{\epsilon_{pn}+Q}})^2.
\label{krakow2}
\end{equation}
We have fitted the function given by Equation~\ref{krakow2} (with C as the only free parameter)
to the data in the excess energy range from 0 to 40~MeV
where the higher partial waves of the nucleon-nucleon system are suppressed~\cite{hab}.
The result is presented in Figure~\ref{pneta-deta}(left) as the solid line, and explains to some extent
the observed decrease of the ratio at threshold. The parameter $C$ was found to 
be $6.85\pm 0.63$ and the $\chi^2$ of the fit was equal to 1.6. 

A slight bump-like structure observed in the ratio presented in the left panel of Figure~\ref{pneta-deta} -- 
with a flat maximum at the excess
energy of about 50~MeV -- could be due to the fact that production of the $\eta$ meson in hadronic 
collisions proceeds via the intermediate resonance N$^*(1535)$
(m(N$^*$)~-~m$_\eta$~-~m$_{nucleon}$~$\approx$~49~MeV).
This may indicate that coupling of this resonance
to the neutron--$\eta$ may be stronger than to the proton-$\eta$ channel.
This
interpretation is however controversial since it would imply 
a strong isospin breaking effects~\cite{kanzo}.

\section{$pn\to pn\eta$(I~=~0) and $pn\to d\eta$ total cross section ratio}

The total cross section for the $pn\to d\eta$ reaction is 
a pure isospin I~=~0 since both deuteron and the $\eta$ meson
have isospin equal to zero.  
In the case of the 
$NN\to NN\eta$ reaction
the I~=~0 component  of the 
cross section
can be extracted from cross sections for the reactions $pn\to pn\eta$  and $pp\to pp\eta$
employing equations~\ref{skladowe} and~\ref{eq3}.

In order to compare 
the production of the $\eta$ meson associated with the proton-neutron bound state~(d$\eta$) 
to its production with the proton and neutron in continuum~(pn$\eta$)
in the way independent of the 
initial state interaction
we have extracted the experimental values of the $\sigma_0(pn\to pn\eta)$ component, 
and compared them to the total cross sections for the $pn\to d\eta$ reaction. 
The ratio of these two cross sections is presented in the right panel 
of Figure~\ref{pneta-deta}, plotted as 
a function of the excess energy Q. 
An interesting observation is this ratio
rises nearly linearly with the excess energy up to circa 60~MeV, and above this 
value it starts to grow more steeply. This may suggest that from about 
60~MeV influence of the higher partial waves in the $pn\eta$ system 
is more pronounced than in the case of the $d\eta$ system.  

According to the reference~\cite{wilkin} 
the low energy cross section for the $pn\to d\eta$ reaction may be parameterized in the following way:
\begin{equation}
\sigma(pn\to d\eta) \approx a \sqrt{Q} (1 + bQ),
\label{deta_cross} 
\end{equation}
where parameters $a$ and $b$ are calculated from the nucleon mass, the $\eta$ and $\rho$ meson masses, and also 
the $\rho$ meson coupling constant~\cite{wilkin}. As the latter is still not well known, and the 
values of parameters $a$ and $b$ are model dependent~\footnote{In reference~\cite{wilkin} they are calculated 
in the framework of the vector meson dominance one boson exchange model, where the $\rho$ meson 
exchange current dominates the production amplitude.} we have treated
$a$ and $b$ as free parameters of the fit. 

Dividing Equation~\ref{faldt} by Equation~\ref{deta_cross} we get: 
\begin{equation}
\frac{\sigma_0(pn\to pn\eta)}{\sigma(pn\to d\eta)} = \frac{DQ^{3/2}}{(1+bQ)(1+\sqrt{1+Q/\epsilon_{pn}})^2},
\label{fit_func}
\end{equation}
with $D$ being a constant. 
We have fitted Formula~\ref{fit_func} in the range between the threshold and 
Q~=~50~MeV 
(see Figure~\ref{pneta-deta} (right))~\footnote{This corresponds to the 
close-to-threshold reaction region, where higher partial waves shouldn't be present, and 
Formulae~\ref{faldt} and~\ref{deta_cross} apply.}, 
treating $D$ and $b$ as free parameters of the fit. The fit procedure resulted in 
$D~=~0.35\pm 0.03~\frac{1}{MeV^{3/2}}$ and $b~=~-0.013\pm 0.001~\frac{1}{MeV}$. 
The value of reduced $\chi^2$ of the fit was equal to 1.4.

On the other hand, the assumption that the $\sigma_0(pn\to pn\eta)$ to 
$\sigma(pn\to d\eta)$ ratio is a linear function of Q in the close-to-threshold
region (up to the excess energy of Q~=~50~MeV): 
\begin{equation}
\frac{\sigma_0(pn\to pn\eta)}{\sigma(pn\to d\eta)} = K Q, 
\label{linear}
\end{equation}
yields $K~=~0.027\pm 0.001~\frac{1}{MeV}$, 
with the reduced value of $\chi^2$ equal to 0.3. 
The best linear function fitted to the experimental data 
is presented in Figure~\ref{pneta-deta} (right). 

One should, however, be careful in interpretation of the 
cross section ratios presented in Figure~\ref{pneta-deta}
due to the rather low 
energy resolution for measurements of the  $pn\to pn\eta$ reaction 
which  was equal to  about 5~MeV for the COSY-11 experiment 
and circa 8~MeV for experiments performed with the WASA/PROMICE detector~\cite{calenpn}. 

\section{Acknowledgements}
The work was partially supported by the
European Community-Research Infrastructure Activity
under the FP6 programme (Hadron Physics,
RII3-CT-2004-506078), by
the Polish Ministry of Science and Higher Education under grants
No. 3240/H03/2006/31  and 1202/DFG/2007/03,
and by the German Research Foundation (DFG).

\end{document}